\documentclass[aps,reprint,letterpaper,showpacs,amsmath,amssymb, superscriptaddress]{revtex4-1}
\usepackage{graphicx}
\usepackage{comment}
\usepackage{color}
\usepackage[absolute]{textpos}

\begin{document}
\begin{textblock*}{8.5in}(0.1in,0.25in)
\begin{center}
PHYSICAL REVIEW B \textbf{92}, 165307 (2015)
\end{center}
\end{textblock*}
\begin{textblock*}{2.5in}(5.06in,3.86in)
{\small \doi{10.1103/PhysRevB.92.165307}}
\end{textblock*}

\title{Electron and hole $g$ factors in InAs/InAlGaAs
self-assembled quantum dots emitting at telecom wavelengths}

\author{V.V. Belykh}
\email[]{vasilii.belykh@tu-dortmund.de}
\affiliation{Experimentelle Physik 2, Technische Universit\"{a}t
Dortmund, D-44221 Dortmund, Germany}
\author{A. Greilich}
\affiliation{Experimentelle Physik 2, Technische Universit\"{a}t
Dortmund, D-44221 Dortmund, Germany}
\author{D.R. Yakovlev}
\affiliation{Experimentelle Physik 2, Technische Universit\"{a}t
Dortmund, D-44221 Dortmund, Germany} \affiliation{Ioffe
Institute, Russian Academy of Sciences, 194021
St. Petersburg, Russia}
\author{M. Yacob}
\affiliation{Institute of Nanostructure Technologies and Analytics (INA), CINSaT, University of Kassel, Heinrich-Plett-Str. 40, D-34132 Kassel, Germany}
\author{J.P. Reithmaier}
\affiliation{Institute of Nanostructure Technologies and Analytics (INA), CINSaT, University of Kassel, Heinrich-Plett-Str. 40, D-34132 Kassel, Germany}
\author{M. Benyoucef}
\affiliation{Institute of Nanostructure Technologies and Analytics (INA), CINSaT, University of Kassel, Heinrich-Plett-Str. 40, D-34132 Kassel, Germany}
\author{M. Bayer}
\affiliation{Experimentelle Physik 2, Technische Universit\"{a}t
Dortmund, D-44221 Dortmund, Germany} \affiliation{Ioffe
Institute, Russian Academy of Sciences, 194021
St. Petersburg, Russia}

\date{\today}
\begin{abstract}
We extend the range of quantum dot (QD) emission energies where
electron and hole $g$ factors have been measured to the practically
important telecom range. The spin dynamics in
InAs/In$_{0.53}$Al$_{0.24}$Ga$_{0.23}$As self-assembled QDs with
emission wavelengths at about 1.6~$\mu$m grown on InP substrate is investigated by
pump-probe Faraday rotation spectroscopy in a magnetic field.
Pronounced oscillations on two different frequencies, corresponding
to the QD electron and hole spin precessions about the field are
observed from which the corresponding $g$ factors are determined.
The electron $g$ factor of about $-1.9$ has the largest negative value so
far measured for III-V QDs by optical methods. This value, as well
as the $g$ factors reported for other III-V QDs, differ from those expected for bulk semiconductors at
the same emission energies, and this difference increases significantly for decreasing energies.
\end{abstract}

\maketitle

\section{Introduction}

Spin physics has attracted great attention in recent years, inspired
by the possibility of using electron or hole spins for storing and
encoding quantum information \cite{Dyakonov2008}. Semiconductor
quantum dots (QDs) provide an appropriate platform for manipulating a
carrier spin, in particular, that of an electron. The spatial
confinement of electrons in QDs suppresses the most efficient spin
relaxation mechanisms \cite{Khaetskii2000} and results in long spin
coherence times \cite{Greilich2006Sci}. One of the most important
parameters for the spin control is the $g$ factor, which
characterizes the susceptibility of a spin to a magnetic field. In
semiconductors, electrons are quasiparticles and their $g$ factor
might be drastically different from the $g_0\approx2$ of a free
electron. $g$ factors in QDs have been measured either electrically
\cite{Medeiros-Ribeiro2002, Medeiros-Ribeiro2003, Potok2003,
Bjork2005, Schroer2011, Alegre2006, Takahashi2013, Prechtel2015} or
optically \cite{Bayer1999, Tischler2002, Goni2000, Nakaoka2004,
Nakaoka2005, Cade2006, Kleemans2009, Greilich2006, Yugova2007,
Dutt2005, Schwan2011, Schwan2011a, Syperek2012, Crooker2010,
Debus2014}. The most widespread optical method is the measurement of
the Zeeman splitting in magnetoluminescence spectra which, in
general, gives only the exciton $g$ factor \cite{Bayer1999,
Tischler2002, Goni2000, Nakaoka2004, Nakaoka2005, Cade2006,
Kleemans2009}. However, in some cases, reduced symmetry of the QDs
has allowed one to observe also the dark exciton states and separate
electron and hole $g$ factor contributions \cite{Bayer1999,
Tischler2002}. Other optical methods of $g$ factor determination are
spin noise spectroscopy \cite{Crooker2010} and spin-flip Raman
scattering \cite{Debus2014}. Especially high precision in the
measurement can be achieved with optical pump-probe spectroscopy,
where the $g$ factor is determined from the frequency of spin
polarization oscillations in a perpendicular magnetic field
\cite{Greilich2006, Yugova2007, Dutt2005, Schwan2011, Schwan2011a,
Syperek2012}. Pump-probe spectroscopy allows one to determine
separately the electron and hole $g$ factors as well as the $g$
factor spread, which contributes to the decay of the oscillations
\cite{Yugova2007}. Furthermore, the spin mode-locking effect in the
pump-probe signal enables one to evaluate the spin coherence time
$T_{2}$ and study the dynamics of the nuclear spin
polarization \cite{Greilich2006Sci, Greilich2007Sci}.

So far, electron and hole $g$ factors have been measured for QDs
emitting in the energy range $E \gtrsim 1.0$~eV. Moreover, in
pump-probe experiments, $g$ factors have been measured only for $E
\gtrsim 1.3$~eV, the energies accessible by a Ti:sapphire laser.
However, an important energy range from a practical point of view is
the telecommunication range which covers $0.75$~eV~$\lesssim E
\lesssim$~0.95~eV ($1.3-1.7$~$\mu$m), corresponding to the
transparency window of an optical fiber. Furthermore, the spin
dynamics in QDs with low band-gap energies is of fundamental
interest as it can be used to test the existing theories of $g$
factors in QDs \cite{Kiselev1998, Rodina2003, Sheng2007, VanBree2012}
and stimulate the development of novel approaches, which in turn can help in refining band structure parameters. In particular,
large in magnitude, negative electron $g$ factors are expected for
small band-gap energies so that they may help to study and
implement new robust spin interaction effects in QDs \cite{Varwig2014,
Varwig2014a}.

\begin{figure*}
\begin{center}
\includegraphics[width=0.55\columnwidth]{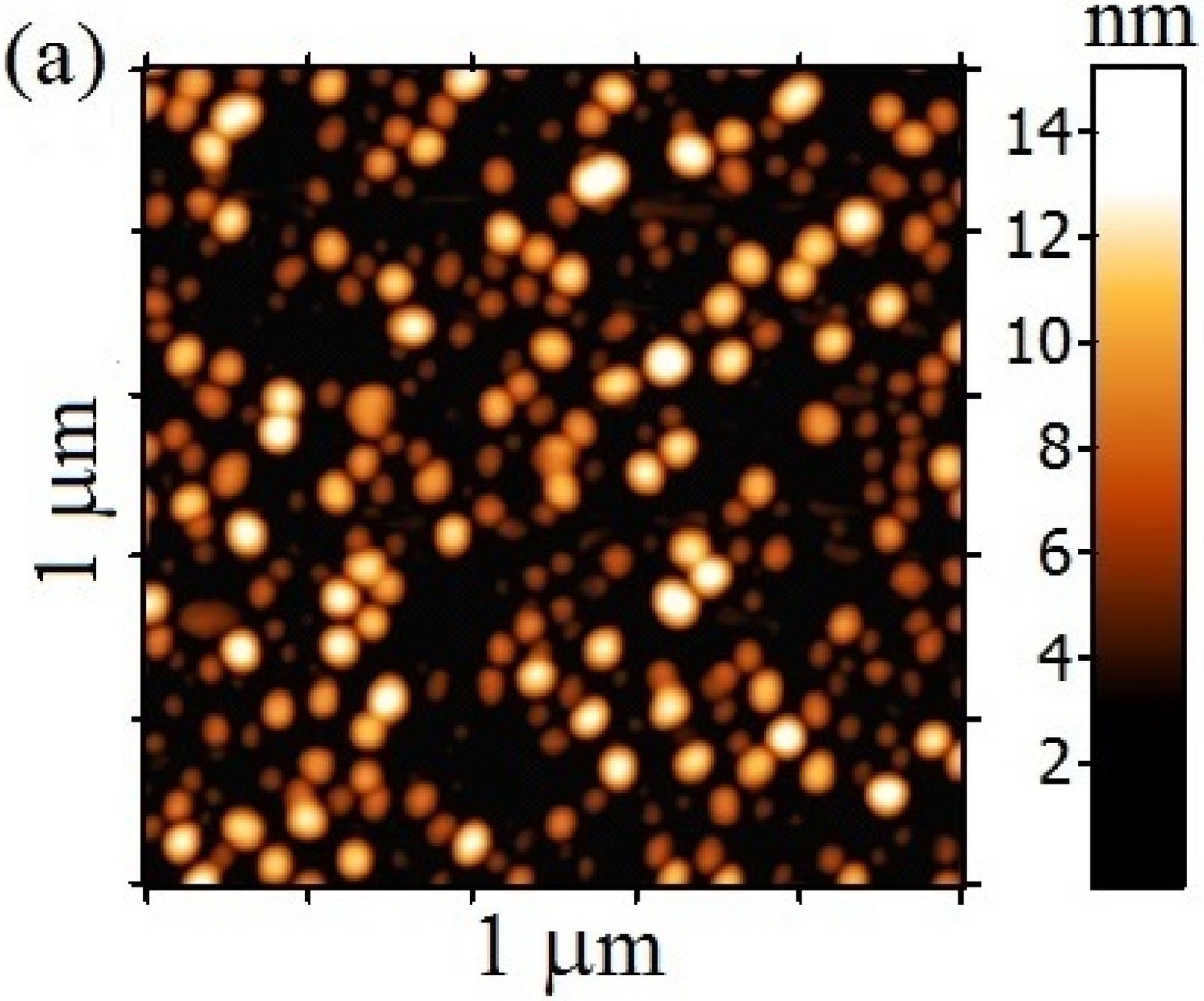}
\includegraphics[width=1.2\columnwidth]{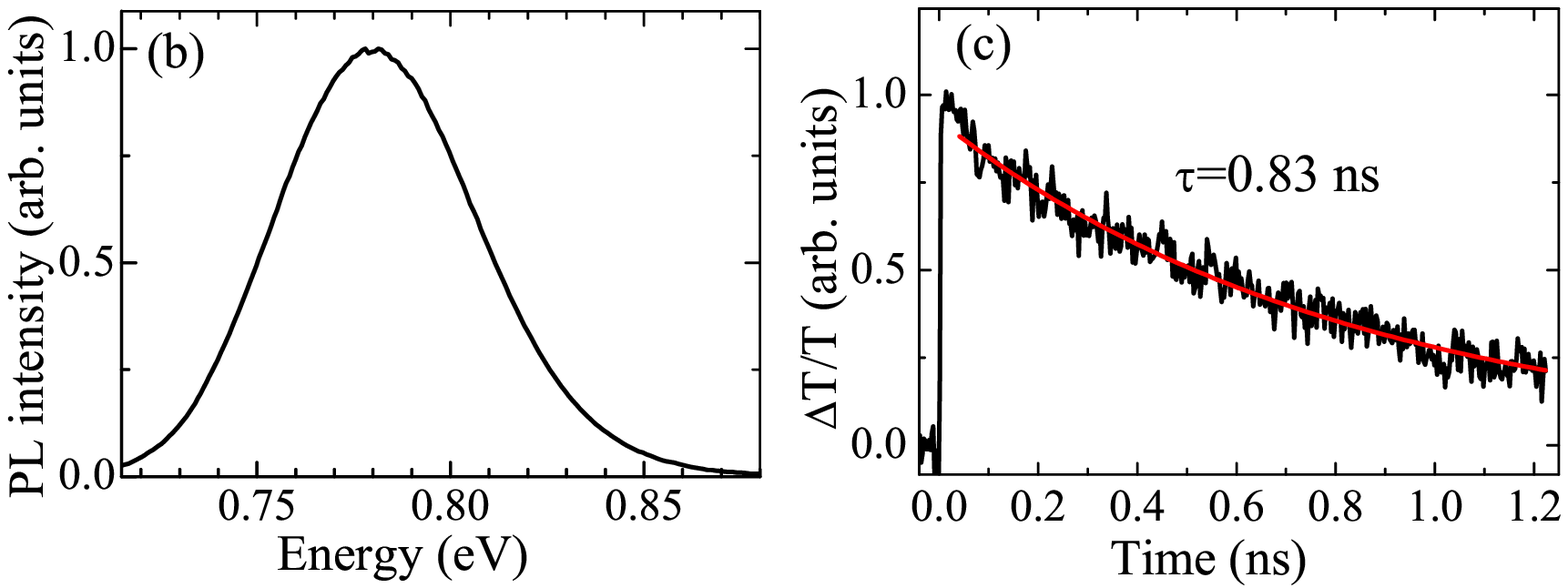}
\caption{(a) Atomic force microscopy image of an InAs QD layer
deposited on an In$_{0.53}$Al$_{0.24}$Ga$_{0.23}$As barrier. (b)
Photoluminescence spectrum of the studied
InAs/In$_{0.53}$Al$_{0.24}$Ga$_{0.23}$As QDs at $T=10$~K. (c) Dynamics of the QDs relative transmission measured with a laser energy of 0.79 eV and spectral width of 10 meV at $T=12$~K. The red line shows the exponential fit to the experimental data.} \label{FigSpec}
\end{center}
\end{figure*}
In this paper we measure the electron and hole $g$ factors, $g_\text{e}$ and $g_\text{h}$, for
QDs emitting around 0.8~eV (1.6~$\mu$m) by using the pump-probe
Faraday ellipticity (analogous to the Faraday rotation) technique.
The obtained electron and hole $g$ factors have the largest absolute
values measured so far for III-V QDs by optical methods. We also
systemize the values of electron $g$ factors
$g_\text{e}^\text{QD}(E)$ for III-V QDs with widely varying emission
energies $E$ and show that $g_\text{e}^\text{QD}(E)
> g_\text{e}^\text{bulk}(E)$, where $g_\text{e}^\text{bulk}(E)$ is the electron
$g$ factor in bulk materials calculated according to the
Roth-Lax-Zwerdling relation \cite{Roth1959} (which gives good
agreement for the electron $g$ factors measured in bulk
semiconductors \cite{Kosaka2001}). The electron $g$ factor in QDs
depends not only on the transition energy $E$, as it is the case for
the longitudinal electron $g$ factor in QWs \cite{Yugova2007Univ},
but is also determined by the QD shape and composition, and can
provide information on these parameters.

\section{Experimental details}
The sample under study was grown by molecular-beam epitaxy on a
(100)-oriented InP substrate and contains 5.5 monolayers of InAs
surrounded by In$_{0.53}$Al$_{0.24}$Ga$_{0.23}$As barriers. The
bottom barrier contains a Si $\delta$-doped layer at a distance of
15~nm from the InAs layer. The InAs layer is transformed into
self-assembled QDs with a density of about $10^{10}$~cm$^{-2}$. An
atomic force microscopy image of the InAs QD layer on top of the
In$_{0.53}$Al$_{0.24}$Ga$_{0.23}$As barrier is shown in
Fig.~\ref{FigSpec}(a). From previous studies of similar structures
it is known that the medium and large sized QDs are optically active,
while the background dots of small size are optically inactive
\cite{Benyoucef2013}. The average diameter and height of the optically active QDs are about 50 and 13~nm, respectively. A photoluminescence (PL) spectrum of such a QD
ensemble taken at temperature $T=10$~K is shown in
Fig.~\ref{FigSpec}(b). The emission is centered at $\sim 0.8$ eV
($\sim1.6$~$\mu$m) with an inhomogeneous broadening originating from
the spread of QD parameters.

The sample is placed in a split-coil magneto-cryostat at $T = 7$~K.
Magnetic fields up to $B=4$~T are applied in the Voigt geometry
(parallel to the sample surface, perpendicular to the light wave
vectors) unless otherwise stated. A pump-probe technique with
polarization sensitivity is employed to measure the spin dynamics. We
use a NT\&C laser system consisting of an optical parametric
amplifier (OPA) pumped by a mode-locked Yb:KGW laser operating at
1040~nm \cite{Krauth2013}. The laser system generates a periodic
train (emission pulse frequency $40$ MHz) of 300-fs-long pulses at a
tunable wavelength of $1350-4500$~nm. By means of a pulse shaper,
the broad ($\sim 60$~nm) spectrum is shaped down to a width of 10~nm (5 meV)
centered at the desired wavelength (1570~nm), unless otherwise
stated. The laser output is split into pump and probe beams. The
circular-polarized pump generates the carrier spin polarization
whose temporal evolution is probed by measuring the ellipticity of
the probe beam, which is initially linearly polarized, after
transmission through the sample. This method is analogous to
measuring the Faraday rotation of the probe beam and provides
similar information \cite{Varwig2012}. In all experiments, except
those where the pump power dependence of the signal strength is
measured, nearly $\pi$-pulse excitation power leading to maximal spin
polarization is used.

The population dynamics of the optically injected electron-hole pairs in the QDs is investigated
by measuring the differential transmission $\Delta T/T$ in a pump-probe experiment (Fig.~\ref{FigSpec}(c)). Linearly polarized pump pulses are used to excite a carrier population that was monitored by the polarized probe pulses as a function of delay relative to the pumps. Pump and probe pulses have orthogonal linear polarizations to avoid  polarization interference and have the same photon energy centered around the maximum of the PL spectrum of the QDs. The obtained dynamics of the transmission (see Fig.~\ref{FigSpec}(c)) shows, to a good approximation, a monoexponential decay corresponding to an exciton recombination time of 0.83~ns.

\section{Results and discussion}
\begin{figure}
\begin{center}
\includegraphics[width=\columnwidth]{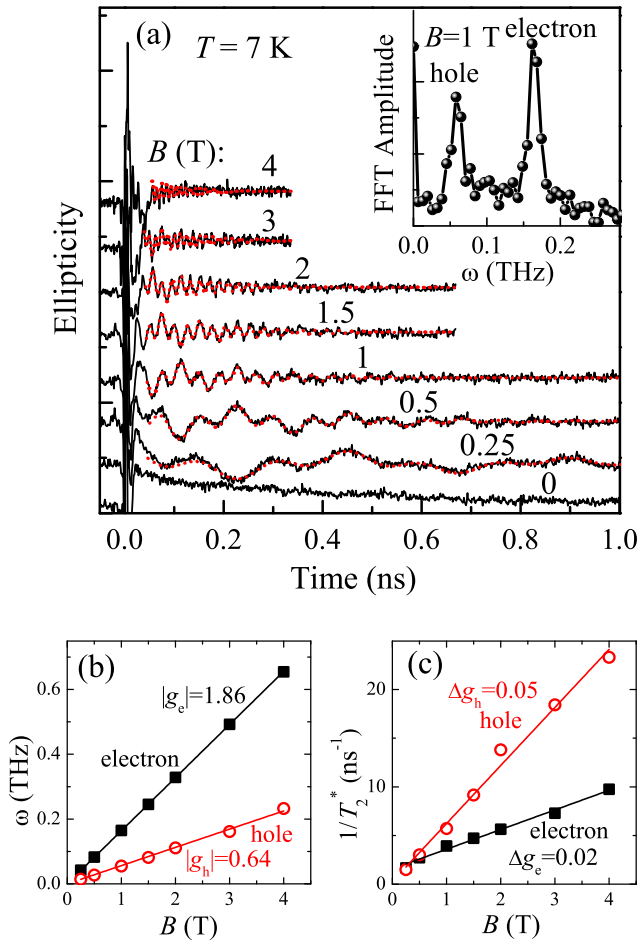}
\caption{(a) Dynamics of the ellipticity signal at different
magnetic fields. The red dotted lines show fits to the experimental
data. The form of the fit is given by the sum of two damped
oscillating functions. The curves are shifted vertically for
clarity. The inset shows a FFT spectrum of the ellipticity signal at
$B=1$~T. (b),(c) Magnetic field dependencies of the oscillation
frequencies (b) and decay rates (c) for the fast (solid squares) and slow (open circles) oscillations in the
spin dynamics. The laser photon energy is set to 0.79 eV.}
\label{FigBDep}
\end{center}
\end{figure}

Figure~\ref{FigBDep}(a) shows the dynamics of the ellipticity signal
recorded on the InAs quantum dot sample for different magnetic
fields. The traces show oscillations on two frequencies as evidenced
by the fast Fourier transform (FFT) spectrum of the ellipticity
dynamics at $B=1$~T shown in the inset of Fig.~\ref{FigBDep}(a). As we
will show below, the fast and slow oscillations can be attributed to
the electron and hole spin precessions, respectively. The dynamics
are fitted by a form representing the sum of two oscillating
functions of type $\cos(\omega t)\exp(-t^2/2T_\text{2}^{*2})$, where
$t$ is the delay time, $\omega$ is the oscillation frequency, and
$T_2^*$ is the spin dephasing time. The fits are shown by the
red dotted lines in Fig.~\ref{FigBDep}(a). The Gaussian type of the
oscillation decay $\exp(-t^2/2T_\text{2}^{*2})$ reflects the
Gaussian spread of a $g$ factor $\exp[-(g-g_0)^2/2\Delta g^{2}]$ \cite{Note0}. This spread is assumed to be the main source of the observed damping of the ellipticity signal in the magnetic field. While the $g$ factor
determines the oscillation frequency $\omega =
|g|\mu_\text{B}B/\hbar$, the spread of the $g$ factor determines the
damping rate $T_\text{2}^{*-1} = \Delta g\mu_\text{B}B/\hbar$,
where $\mu_\text{B}$ is the Bohr magneton. The other possible source
of the oscillation damping is fluctuations in the nuclear spin bath
in the quantum dot \cite{Merkulov2002}. Their contribution, however,
shows up only for magnetic fields much weaker than the $B$
considered here. Note that oscillations are observed only within the carrier lifetime ($\approx0.83$~ns) which also limits the spin coherent signal in QDs without resident carriers (uncharged QDs).

The dependencies of the oscillation frequencies on the magnetic
field are shown in Fig.~\ref{FigBDep}(b). They give the following
values of the transverse $g$ factors: $|g_\text{e}| = 1.86$ for the fast
(electron) oscillations and $|g_\text{h}| = 0.64$ for the slow
(hole) oscillations. Figure~\ref{FigBDep}(c) shows the dependencies
of the damping rates $T_\text{2}^{*-1}$ on the magnetic field. They
are close to linear, confirming that the $g$ factor spread is the
main source of the spin dephasing. Linear fits to the measured
dependencies give the following values of the $g$ factor spreads:
$\Delta g_\text{e} = 0.02$ for the fast oscillations and $\Delta
g_\text{h} = 0.05$ for the slow oscillations. The nonzero offset of the linear dependencies is related to other spin dephasing mechanisms as well as to the exciton recombination.

The value $|g_\text{e}| = 1.86$ for the fast oscillations is larger
than the moduli of the electron $g$ factors so far measured for QDs with emission at higher energies \cite{Greilich2006Sci, Medeiros-Ribeiro2002,
Medeiros-Ribeiro2003, Bayer1999, Tischler2002, Greilich2006,
Yugova2007, Dutt2005, Schwan2011, Schwan2011a, Crooker2010,
Debus2014}. We attribute the fast oscillations to
electron spin precession. However, $|g_\text{e}| = 1.86$ is
smaller than the value of $|g_\text{e}| \approx 5$ that one would
expect for $E_\text{g}=0.79$~eV from the Roth-Lax-Zwerdling relation
for bulk semiconductors \cite{Roth1959},
\begin{equation}
g_\text{e}(E_\text{g}) = g_\text{0} - \frac{2E_\text{p}
\Delta_\text{SO}}{3E_\text{g}(E_\text{g}+\Delta_\text{SO})},
\label{EqRLZ}
\end{equation}
where $E_\text{g}$ is the band-gap energy,
$\Delta_\text{SO}$ is the spin-orbit splitting of the valence band,
and $E_\text{p}=2P_\text{cv}^2/m_0$ is the Kane energy ($P_\text{cv}$ is the interband momentum matrix element and $m_0$ is the free electron mass).

The origin of the slow oscillations is less evident. It may be
attributed to the hole spin precession. However, the measured value
of $|g_\text{h}| = 0.64$ is much larger than the transverse hole $g$
factor measured for annealed (In,Ga)As/GaAs QDs emitting around
1.4~eV ($|g_{\text{h}\perp}| \sim 0.2$) \cite{Schwan2011}. An other possible origin
of the slow oscillations might be electron spin precession in the
InP substrate or in the In$_{0.53}$Al$_{0.24}$Ga$_{0.23}$As barriers which may be initiated through
two-photon absorption. A similar situation was described in
Ref.~\cite{Yugova2007Univ}, where electron spin precession in the
GaAs buffer is superimposed on the signal from GaAs/(Al,Ga)As
quantum wells.

\begin{figure}
\begin{center}
\includegraphics[width=\columnwidth]{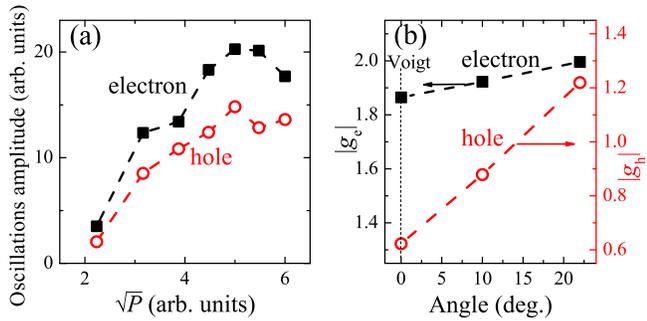}
\caption{(a) Dependence of the oscillation amplitudes on the square
root of the pump power for the fast (solid squares) and slow (open
circles) oscillations in the observed spin precessions. Broadband
$\sim30$ meV pump and probe beams without a pulse shaper are used. (b)
Dependence of the $g$ factor moduli on the angle between the sample
surface and the magnetic field for the fast (solid squares, left
axis) and slow (open circles, right axis) oscillations. The zero angle
corresponds to the Voigt geometry. (a),(b) The dashed lines are guides
to the eye. $B=1$~T, the laser photon energy is 0.79 eV, and $T=7$~K.}
\label{FigPADep}
\end{center}
\end{figure}

To clarify the origin of the slow oscillations, we measured the pump
power dependence of the oscillation amplitudes at $B=1$~T. The pump
power $P$ defines the pump pulse area $\propto \int E(t) dt \propto
\sqrt{P}$, where $E(t)$ is the electric field amplitude \cite{Scully1997}.
Figure~\ref{FigPADep}(a) shows the dependence of the oscillation
amplitude on $\sqrt{P}$ for the fast and slow oscillations. Both
amplitudes have a pronounced maximum, presumably corresponding to the
pulse area of $\pi$. The presence of a maximum in
the dependencies indicates that the corresponding oscillations are
related to the Rabi oscillations in the QD excitation, while in bulk
such oscillations are hard to observe due to the fast excitation
induced dephasing for elevated excitation power. Furthermore, the similar
behavior for both dependencies suggests a common source for both
precession frequencies. These facts exclude a barrier/substrate origin of the
slow oscillations.

To further confirm the hole nature of the slow oscillations, we
measured the spin dynamics for nonzero angles of the magnetic field
relative to the sample surface at $B=1$~T. It is well known that the
hole $g$ factor is strongly anisotropic in (In,Ga)As QDs
\cite{Schwan2011} and can be several times higher for the magnetic
field parallel to the sample growth axis than in transverse
magnetic field. On the other hand, the electron $g$ factor is more
isotropic, which allows one to distinguish electron and hole spin
beats. Indeed, tilting the sample with respect to the magnetic field
by an angle of $\sim 20^o$ leads to a slight increase of $|g_\text{e}|$
by $\sim 0.1$ (Fig.~\ref{FigPADep}(b), left axis) compared to a
significant increase of $|g_\text{h}|$ by $\sim 0.6$
(Fig.~\ref{FigPADep}(b), right
axis).
The large hole $g$ factor in the studied unannealed QDs can be
explained by admixing the light-hole states to the heavy-hole states as
a result of strong spatial confinement.

\begin{figure}
\begin{center}
\includegraphics[width=\columnwidth]{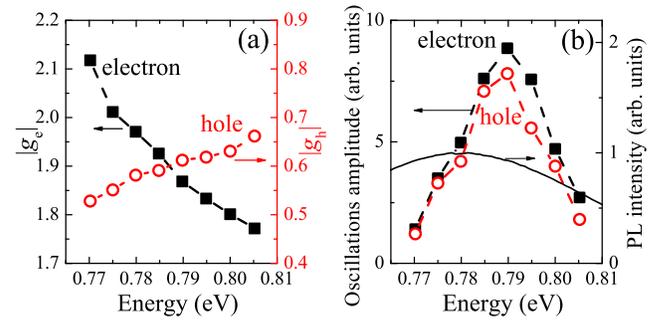}
\caption{Dependence of $g$ factor moduli (a) and oscillation
amplitudes (b) on the laser energy for the electron (solid squares) and
hole (open circles) spin precession signal. The solid line shows the
PL spectrum (right axis). (a),(b) The dashed lines are guides to the
eye. $B=1$~T, $T=7$~K.} \label{FigEDep}
\end{center}
\end{figure}

It is interesting to examine how the electron and hole $g$ factors
are affected by the spread of QD parameters within an ensemble. The
spread of QD parameters manifests itself in the inhomogeneous
broadening of the PL spectrum (Fig.~\ref{FigSpec}(b)). We studied the
spin precessions at $B=1$~T as functions of the laser photon energy (the laser spectral width is $\sim 5$ meV)
which selects certain QD subsets in the whole ensemble. The energy
dependence of the electron and hole $g$ factor moduli are shown in
Fig.~\ref{FigEDep}(a) by the solid squares (left axis) and open
circles (right axis), respectively. Interestingly, the modulus of
the electron $g$ factor decreases with energy, as reported for the
electron $g$ factor in (In,Ga)As/GaAs QDs \cite{Schwan2011a,
Debus2014, Greilich2006Sci} and GaAs/(Al,Ga)As QWs
\cite{Yugova2007Univ} and expected from Eq.~\eqref{EqRLZ} for
negative $g$ factors. The negative sign of the electron $g$ factor
was also proven previously for (In,Ga)As/GaAs QDs emitting at larger
energies by measuring the dynamic nuclear polarization
\cite{Yugova2007}. On the other hand, the modulus of the hole $g$
factor increases with energy. Such a behavior has been reported only
for holes \cite{Crooker2010}.

The emission energy dependencies of the amplitudes of the electron
and hole oscillations (Fig.~\ref{FigEDep}(b)) show a similar peaked
behavior with the maximum close to the PL maximum energy (solid line in
Fig.~\ref{FigEDep}(b), corresponding to the right axis), which
further confirms the QD origin of both oscillations. However, the
width of the dependencies ($\sim 20$~meV) is several times smaller
than the width of the PL spectrum ($\sim 60$~meV). The comparable amplitudes of the electron and hole spin precessions indicate almost equal electron and hole populations. This fact, together with the observed oscillation decay times not exceeding the carrier population decay time ($\approx0.83$~ns), suggest that the concentration of charged QDs is low and empty QDs dominate the signal despite the Si $\delta$-doping layer.

\begin{figure}
\begin{center}
\includegraphics[width=\columnwidth]{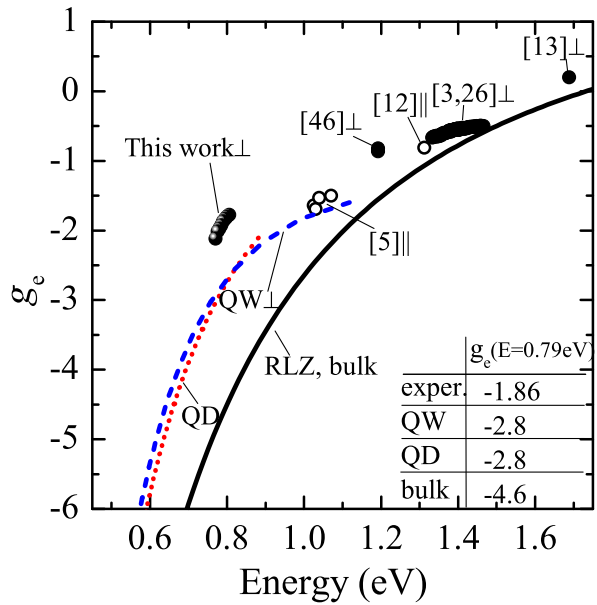}
\caption{Dependence of the electron $g$ factor on the QD emission
energy. The solid and open symbols correspond to the transverse and
longitudinal $g$ factors, respectively. The original data from the
present work are shown together with the data from
Refs.~\cite{Medeiros-Ribeiro2003, Bayer1999,
Tischler2002, Greilich2006Sci, Debus2014, Note1}. The solid line
shows the dependence for bulk semiconductors calculated according to
the Roth-Lax-Zwerdling relation~\eqref{EqRLZ}. The dotted and dashed
lines show the dependencies calculated for spherical QDs and QWs
using the model of Ref.~\cite{Kiselev1998}. The table shows the $g$
factor values determined experimentally and calculated using
different approaches at $E=0.79$~eV.} \label{FigRLZ}
\end{center}
\end{figure}

Figure~\ref{FigRLZ} summarizes the electron $g$ factors obtained in
the present work, together with the data reported in literature for
III-V QDs \cite{Medeiros-Ribeiro2003,
Bayer1999, Tischler2002, Greilich2006Sci, Debus2014,
Note1}, as a function of the QD emission energy. The solid symbols
correspond to the transverse $g$ factors ($\mathbf{B}$ is
perpendicular to the growth axis), while open symbols correspond to
the longitudinal $g$ factors ($\mathbf{B}$ is parallel to the growth
axis). The solid line shows the $g$ factor energy dependence
$g_\text{e}^\text{bulk}(E)$ calculated for bulk semiconductors
according to Eq.~\eqref{EqRLZ} \cite{Roth1959}. From
Fig.~\ref{FigRLZ} it is clear that $g_\text{e}^\text{QD}(E) >
g_\text{e}^\text{bulk}(E)$ without any exception. The deviation
between QD and bulk $g$ factors is maximal for the lowest emission energies.
This deviation is presumably related to the effect of confinement on
the spin-orbit coupling.

To account for the confinement effect, we calculated the energy
dependence of the electron $g$ factor using the theory of
Ref.~\cite{Kiselev1998}, which is based on the Kane's model. We use
two different approaches for the QD shape: (i) a spherical QD and
(ii) a flattened QD approximated by a QW. The second approach is
more realistic for the studied QDs with a dome shape. The QD
material is assumed to be InAs with the band-gap energy $E_\text{g}
= 0.417$~eV, the spin-orbit splitting of the valence band
$\Delta_\text{SO} = 0.39$~eV, the interband matrix element
$E_\text{p} = 21.5$~eV, and the heavy-hole mass $m_\text{hh}=0.45m_0$
\cite{Vurgaftman2001, Adachi1992}. A conduction- to valence-band
offset ratio of $\Delta E_\text{c}/\Delta E_\text{v}=0.6/0.4$ is
used. For the In$_{0.53}$Al$_{0.24}$Ga$_{0.23}$As barriers we used
intermediate parameters between those of InAs and GaAs which are
determined for the band-gap energy $E_\text{g} = 1.2$~eV by linear
interpolation between the InAs and GaAs parameters as a function of
the band-gap energy: $\Delta_\text{SO} = 0.36$~eV, $E_\text{p} =
26.7$~eV, $m_\text{hh}=0.45m_0$. Linear interpolation between the InAs
and GaAs parameters was also used to determine $\Delta_\text{SO}$
and $E_\text{p}$ as a function of energy for calculating
$g_\text{e}^\text{bulk}(E)$ according to Eq.~\eqref{EqRLZ} (the
solid line in Fig.~\ref{FigRLZ}). In all cases we added
$g_\text{remote}=-0.13$ to the calculated $g$ factors to account for
the contribution from the remote bands \cite{Kiselev1998,
Yugova2007Univ} not included directly in the calculation. Note that
for the bulk case and for spherical QDs, the longitudinal components
of the $g$ factors coincide with the transverse components.

The results of the calculations in the QD approach are shown in
Fig.~\ref{FigRLZ} by the dotted line and the calculated dependence
for the QW approach is shown by the dashed line. Different energies
for the calculated dependencies correspond to different QD radii in
the QD approach and different QW widths in the QW approach. The
dependencies are much closer to the experimental values than that
according to Eq.~\eqref{EqRLZ}, but a significant deviation still
remains. The results are summarized in the table shown in the inset
of Fig.~\ref{FigRLZ}. It is worth noting that the slope of the
measured dependence around 0.8~eV is close to the slopes of the
dependencies calculated for bulk and within the QD approach.

More realistic calculations of QD electron $g$ factors should take
into account the strain effects which might be significant for
QDs emitting at energies below 1.2~eV \cite{Kiselev1999}. These effects can
lead to renormalization of the QD band gap and induce significant
mixing of light-hole and heavy-hole states, which in turn will
change the electron $g$ factor.

\section{Conclusion}
Using pump-probe spectroscopy we studied the spin dynamics in
InAs/In$_{0.53}$Al$_{0.24}$Ga$_{0.23}$As self-assembled quantum dots
emitting in the telecom spectral range around 1.6~$\mu$m.
Oscillations at frequencies corresponding to the transverse $g$
factors $|g_\text{e}|\approx1.9$ and $|g_\text{h}|\approx0.6$ were
observed in the ellipticity signal in a magnetic field and identified
as electron and hole spin beats, respectively. The electron $g$
factor values measured in the present work and reported previously
for III-V QDs are higher than the $g$ factors calculated for bulk
semiconductors using the Roth-Lax-Zwerdling relation \cite{Roth1959}
at the same energy (note the negative sign of the $g$ factors). The discrepancy from the Roth-Lax-Zwerdling relation
increases with decreasing QD emission energy. Calculations within the
Kane's model taking into account the confinement effect
\cite{Kiselev1998}, partly reduce the discrepancy, however, an even
more refined theoretical description of the experimental findings is
still needed.

\begin{acknowledgements}
We are grateful to E.~L.~Ivchenko and A.~A.~Kiselev for valuable
advice and useful discussions and to F.~Heisterkamp and A.~Steinmann for technical support. We acknowledge the financial support
by the Deutsche Forschungsgemeinschaft and the Russian Foundation of
Basic Research in the frame of the ICRC TRR 160 as well as by the
BMBF in the frame of the project Q.com-H (Contracts No. 16KIS0112 and No. 16KIS0104K). M.~Bayer acknowledges support by the
Government of Russia (Project No. 14.Z50.31.0021).
\end{acknowledgements}

\end{document}